\documentclass{article}

\usepackage{arxiv}

\usepackage[utf8]{inputenc} 
\usepackage[T1]{fontenc}    
\usepackage{hyperref}       
\usepackage{url}            
\usepackage{booktabs}       
\usepackage{amsfonts}       
\usepackage{nicefrac}       
\usepackage{microtype}      
\usepackage{lipsum}		
\usepackage{graphicx}
\usepackage{natbib}
\usepackage{doi}

\usepackage{amsmath,amssymb,amsfonts}
\usepackage{algorithmic}
\usepackage{graphicx}
\usepackage{textcomp}
\usepackage{booktabs}                  
\usepackage{bm}

\title{Effects of Multisensory Feedback on the Perception and Performance of Virtual Reality Hand-Retargeted Interaction}

\author{ 
\href{https://orcid.org/0000-0002-6824-3514}{\includegraphics[scale=0.06]{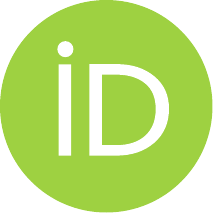}\hspace{1mm}Hyunyoung Jang}\thanks{These authors contributed equally to this work.} \\
	Graduate School of Culture Technology\\
	Korea Advanced Institute of Science and Technology\\
	Daejeon, Republic of Korea \\
	\texttt{jhy439@gmail.com} \\
	\And
	\href{https://orcid.org/0000-0002-1962-5815}{\includegraphics[scale=0.06]{orcid.pdf}\hspace{1mm}Jinwook Kim}\thanks{These authors contributed equally to this work.} \\
	Graduate School of Culture Technology\\
	Korea Advanced Institute of Science and Technology\\
	Daejeon, Republic of Korea \\
	\texttt{jinwook.kim31@kaist.ac.kr} \\
     \And
	\href{https://orcid.org/0000-0002-3403-8117}{\includegraphics[scale=0.06]{orcid.pdf}\hspace{1mm}Jeongmi Lee}\thanks{Corresponding Author} \\
	Graduate School of Culture Technology\\
	Korea Advanced Institute of Science and Technology\\
	Daejeon, Republic of Korea \\
	\texttt{jeongmi@kaist.ac.kr} \\
}

\renewcommand{\shorttitle}{ }

\hypersetup{
pdftitle={Effects of Multisensory Feedback on the Perception and Performance of Virtual Reality Hand-Retargeted Interaction},
pdfsubject={cs},
pdfauthor={Hyunyoung Jang, Jinwook Kim, Jeongmi Lee},
pdfkeywords={Hand-retargeting, Haptic Feedback, Interaction, Multisensory, Perception, Virtual Reality},
}

\begin{document}
\maketitle

\begin{abstract}
	Retargeting methods that modify the visual representation of real movements have been widely used to expand the interaction space and create engaging virtual reality experiences. For optimal user experience and performance, it is essential to specify the perception of retargeting and utilize the appropriate range of modification parameters. However, previous studies mostly concentrated on whether users perceived the target sense or not and rarely examined the perceptual accuracy and sensitivity to retargeting. Moreover, it is unknown how the perception and performance in hand-retargeted interactions are influenced by multisensory feedback. In this study, we used rigorous psychophysical methods to specify users' perceptual accuracy and sensitivity to hand-retargeting and provide acceptable ranges of retargeting parameters. We also presented different multisensory feedback simultaneously with the retargeting to probe its effect on users' perception and task performance. The experimental results showed that providing continuous multisensory feedback, proportionate to the distance between the virtual hand and the targeted destination, heightened the accuracy of users' perception of hand retargeting without altering their perceptual sensitivity. Furthermore, the utilization of multisensory feedback considerably improved the precision of task performance, particularly at lower gain factors. Based on these findings, we propose design guidelines and potential applications of VR hand-retargeted interactions and multisensory feedback for optimal user experience and performance.
\end{abstract}


\section{Introduction}
Recent advanced Virtual Reality (VR) Head Mounted Displays (HMDs) are portable and provide a more comfortable experience. Along with this improvement, various research is ongoing to augment the real-world haptic sense (e.g., weight, stiffness, and resistance) in the virtual environment~\cite{choi2017grabity, je2019aero, lopes2017providing}. Overall, there are two main approaches in this VR haptic research: mechanical and perceptional. First, the mechanical approach utilizes actuators such as motors or vibrators to augment the target sense \cite{whitmire2018haptic, han2020mouille, heo2018thor}. For instance, \cite{lee2019torc} developed a device based on vibrotactile motors and sensors to augment virtual object texture and compliance. On the other hand, the perceptional approach utilizes the illusional aspects that arise from VR HMD's visual information separated from the real world \cite{gonzalez2023sensorimotor, lecuyer2001boundary}. The most widely used methods based on visual illusion are pseudo-haptic feedback \cite{samad2019pseudo, lecuyer2009simulating, kim2021effect} and hand retargeting \cite{zenner2019estimating, cheng2017sparse, han2018evaluating}. These methods manipulate the virtual body to move more or less than the real body movement to trigger the illusion of target sense or expand the interaction space \cite{azmandian2016haptic, zenner2021hart}. Since the perceptional approach is based on modifying the visual representation of real movements, if the offset applied is too large, users could notice the inconsistency between the visual and proprioceptive information, which would seriously hinder immersion and task performance in VR \cite{gonzalez2023sensorimotor, abtahi2022beyond, feuchtner2017extending}. Therefore, it is essential to understand the perceptual factors of illusion-based haptic feedback and specify the range of stimulation that users cannot detect the inconsistency for appropriate usage. 

Important perceptual factors that should be considered include whether the target sense has been enhanced or not, users' sensitivity to changes in stimuli, and the impact on embodiment and presence levels \cite{clark2023effect, azmandian2016haptic}. Particularly, hand retargeting methods that involve continuous modification of movements require accurate characterization of users' perceptual features, such as the psychometric function, the point of subjective equality (PSE), and the threshold range \cite{gonzalez2023sensorimotor, kohm2022sensitivity}. The PSE represents a point at which the comparison stimulus is judged as higher or lower than the standard stimulus with a 50\% probability~\cite{guilford1954psychometric, bock1968measurement, gescheider2013psychophysics}.  The threshold range indicates the range in which users cannot reliably detect changes in stimulation, calculated by the difference between the upper and lower thresholds (i.e., where participants' responses are 75\% and 25\% correct). However, previous studies have primarily focused on whether users perceived the target sense or not, but seldomly examined users' perceptual accuracy and sensitivity to retargeting~\cite{longo2008embodiment, slater2010first, normand2011multisensory, gonzalez2018avatar}. Here, the perceptual accuracy denotes how veridical users' perception is, reflected by the offset of the PSE from the standard point (no retargeting) on the psychometric function; the smaller the offset is, the more accurate the perception is. Also, perceptual sensitivity denotes the ability to discriminate the changes in the retargeting parameter (gain factor), reflected by the slope of the psychometric function and the threshold range; the steeper the slope is, the narrower the threshold range becomes, which reflects higher perceptual sensitivity. For optimal user experience and performance, it is critical to precisely specify users' perception (accuracy and sensitivity) of retargeting and utilize the range of parameters in which users cannot notice the modification.

In addition, research on multisensory feedback is being actively conducted to provide more realistic VR experiences, and it has been confirmed that senses other than vision could also affect augmented haptic perception~\cite{kim2022effect, achibet2015elastic, melo2022multisensory}. These results suggest that combining feedback from multiple sensory modalities might affect users' perception of hand retargeting. For instance, there have been attempts to utilize motors to augment the stretched sense to extend the affordable manipulation range for hand retargeting \cite{yamashita2018gum}. However, those studies lacked a detailed analysis of users' perceptual features \cite{abtahi2022beyond}. Moreover, it is yet to be explored how the perceptual accuracy and sensitivity to hand retargeting are influenced by concurrent multisensory feedback. To design more effective hand retargeting methods for future VR applications, it is necessary to understand the impact of multisensory feedback on human perception and performance in VR.

Thus, this study aims to specify the perceptual features (psychometric function, PSE, and threshold range) for hand retargeting and verify how users' perception and performance are changed by concurrent multisensory feedback. In the experiment, a simple hand-reaching task was performed inside VR while the location of the virtual hand was remapped, and participants responded whether they detected the retargeting or not with the two-alternative forced-choice (2AFC) paradigm. Importantly, continuous sensory feedback (none, visual, visual+auditory, visual+tactile, and visual+auditory+tactile) proportional to the distance between the virtual hand and the reaching destination was provided to observe its impact on users' perceptual accuracy and sensitivity to retargeting as well as task performance. First, we verify the effect of multisensory feedback on the perception of retargeting by deriving the psychometric function of each experimental condition to obtain the offset of the PSE from the standard point (accuracy) and the threshold range (sensitivity). The no-feedback condition is compared with the visual feedback condition, and then, the unimodal feedback (visual) condition is compared with bimodal feedback (visual+auditory, visual+tactile) and trimodal feedback (visual+auditory+tactile) conditions. Second, we verify the effect of concurrent multisensory feedback on task performance by comparing the precision error and task completion time between experimental conditions. Based on the results, we propose that our study makes the following contributions:

\begin{itemize}
\item Using rigorous psychophysical methods, we accurately specify users' perceptual accuracy and sensitivity for hand-retargeting and provide the range of retargeting parameters for optimal user experience and performance in VR applications.  
\item We made a novel attempt to examine how users' perceptions and task performance are influenced by concurrently provided multisensory feedback. Based on the experimental results, we suggest the points to be considered in the design of hand-retargeted interactions that incorporate the changes in perception and task performance that occur with multisensory feedback.
\end{itemize}

\section{Related work}
\subsection{Perceptual Features in Virtual Reality}
Perceptual sensitivity has often been investigated with pain and temperature stimuli in relation to the level of body ownership towards the virtual body.  For example, one study measured the threshold for pressure pain by an electronic pressure algometer, revealing that the feeling of body ownership in the virtual body increased the pressure pain threshold in the real body~\cite{hansel2011seeing}. Another study on temperature perception showed that bodily illusion decreased the sensitivity to temperature stimuli \cite{llobera2013relationship}. These results indicate that the body ownership illusion made people more tolerant of physical stimulation of the real body and decreased perceptual sensitivity. In contrast, a study that used the subjective Likert scale to measure the sensitivity to pain showed the highest sensitivity to the realistic hand, and the sensitivity reduced as the realism of the virtual hand decreased~\cite{lin2016need}. In addition, in Martini and colleagues' work~\cite{martini2015body}, the transparency of the virtual body reduced the sense of embodiment but did not increase pain sensitivity. The inconsistent results regarding the relationship between body ownership and perceptions of stimulation in virtual environments suggest the need for examining how and whether the embodiment level influences the perception of hand retargeting. Thus, we included an embodiment questionnaire in our study and compared the results between experimental conditions.

There have been a few studies that investigated perceptual features in hand remapping techniques. For example, one study examined discrepancy detection thresholds in hand speed~\cite{burns2006perceptual}, and another attempted to use cutaneous haptic feedback to increase detection thresholds~\cite{lee2015enlarging}. However, these two studies utilized a single value, the PSE, to characterize users' perception, disregarding the slope of the psychometric function and the threshold range. This could lead to an inaccurate approximation of perceptual features, as the threshold range is essential for designing acceptable retargeting interaction with the multisensory feedback~\cite{kim2022effect, samad2019pseudo}. Hence, we used both the offset of the PSE from the standard point (accuracy) and the threshold range (sensitivity) for more precise analyses of users' perceptual features. In this study, we investigated how multisensory feedback induces changes in users' perceptual accuracy and sensitivity for hand retargeting.

\subsection{Multimodal Feedback in Virtual Reality}
Multisensory or multimodal feedback in a virtual environment enriches user experience in various aspects. Several studies have proven that multimodal feedback outperforms bimodal and unimodal feedback in terms of task performance and sense of presence~\cite{kim2022effect, dinh1999evaluating, diaz2006influence, hecht2006multimodal, frohlich2013visual, sella2014natural, cooper2018effects}. For example, visual, auditory, and haptic feedback combined together in VR reduced the task completion time for a wheel change task \cite{cooper2018effects}, helped users to avoid undesirable collisions \cite{diaz2006influence}, enhanced memory for objects in the virtual environment \cite{dinh1999evaluating}, speeded mental processing time~\cite{hecht2006multimodal}, improved the performance in a button-pressing task combined with a memory test~\cite{viciana2010influence}, enhanced the performance in a simulated drilling task~\cite{hu2012impact}, and increased the sense of presence \cite{dinh1999evaluating, viciana2010influence, hu2012impact, kim2014design, cooper2018effects}. The addition of haptic feedback alone enhanced paddle juggling performance by reducing variability \cite{ankarali2014haptic}, and decreased the time taken to perform a pointing task \cite{corbett2016effects}. For other studies, adding tactile feedback was proven to be better than adding auditory feedback in improving the performance in various tasks: a reaching task \cite{durlach2005effect}, a collision-avoiding task \cite{diaz2006influence}, and a texture identification task \cite{lederman2003relative}.

Although multimodal feedback often induces positive effects on VR experience and performance, it could also vary depending on the context. For example, auditory and haptic force feedback did not significantly aid selection task performance in a densely occluded virtual environment \cite{vanacken2009multimodal}. Similarly, the presence of haptic feedback did not elicit any effects on reaching performance, and visual feedback alone was enough for high accuracy in reaching motion \cite{ebrahimi2016empirical}. 

In contrast to the extensive work on investigating the effects of multimodal feedback on task performance and sense of presence, its effects on the sense of embodiment or perceptual features are under-explored. A recent review paper revealed that only a few research papers examined the extent to which multisensory stimuli affect the sense of embodiment \cite{melo2022multisensory}. Evidence proves that synchronous multisensory stimulation elicits body ownership illusion \cite{normand2011multisensory}, even with the visuomotor inconsistencies \cite{caola2018bodily}. Other previous studies revealed that visuotactile feedback elicits significantly stronger body ownership illusion than visuoauditory feedback \cite{choi2016multisensory}, and even enhances both body ownership and agency of virtual wings \cite{egeberg2016extending}. The visuotactile feedback enhanced the embodiment of a virtual prosthesis and thus improved control of the prosthesis \cite{cavalcante2021importance}. In addition, several previous studies showed that multisensory feedback supports participants to perceive the target sense more sensitively~\cite{kim2022effect, diaz2006influence, hecht2006multimodal}. Further research is needed to broaden the knowledge about the complex effects of multimodal feedback on the sense of embodiment, perceptual features, and task performance in VR applications.

\begin{figure*}[h]
    \vspace{1ex}
    \centerline{\includegraphics[width=1\linewidth]{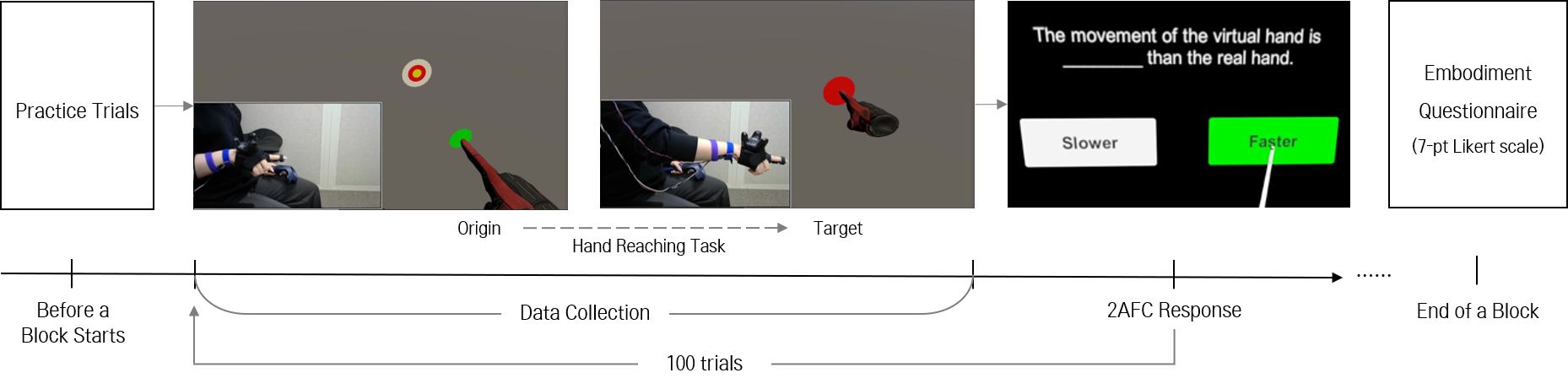}}
    \caption[overall]{The overall experiment procedure. Before a block of each experimental condition starts, participants go through the practice phase. During the experiment, participants perform a simple hand-reaching task while the hand is retargeted and answer the 2AFC question (slower or faster) regarding the speed of the virtual hand's movement compared to the real hand. Each block consists of 100 trials (20 repetitions for five gain factors), and participants respond to the embodiment questionnaire at the end of each block. This procedure is repeated for five experimental conditions with different multisensory feedback.
    } \label{teaser}
\end{figure*}

\section{Method}
\subsection{Multisensory Feedback \& Hand Retargeting Design}
Participants performed a VR hand-reaching task while the location of the virtual hand was retargeted according to one of five gain factors, and responded whether the virtual hand moved faster or slower than the real hand with the 2AFC paradigm \cite{kim2022effect, samad2019pseudo}. Importantly, continuous sensory feedback proportional to the distance between the virtual hand and the reaching destination was provided to observe its impact on perceptual features of hand-retargeting. There were five feedback conditions in our within-subjects experiment:

\begin{enumerate}
    \item No feedback (\textit{None})
    \item Visual feedback (\textit{V})
    \item Visual and auditory feedback (\textit{VA})
    \item Visual and tactile feedback (\textit{VT})
    \item Visual, auditory, and tactile feedback (\textit{VAT})
\end{enumerate}

In the conditions with sensory feedback, stimulation in each sensory modality gradually changed its feature as the virtual hand got closer to the destination (target)  in the following way: 

\begin{itemize}
\item Visual - the edge of the target changed its color from white to red
\item Auditory - the pitch of the auditory signal got higher
\item Tactile - the frequency of vibration got higher
\end{itemize} 
 
In terms of implementing hand retargeting, we used the algorithm developed by~\cite{zenner2019estimating}. 
As a participant touches the origin point and moves toward the target point, the algorithm computes the distance vector ($\Vec{d_r}$) between the physical hand position ($\vec{p_r}$) and the origin point ($\vec{o}$). This distance vector represents the participant's physical reaching movement or the actual position of the real hand from the origin point. 
\[ \Vec{d}_{r} = \vec{p}_{r} - \vec{o} \]
Based on the movements of the physical hand, the virtual hand movements are remapped. From the actual position from the origin ($\Vec{d_r}$), a gain factor (\textit{g}) is applied to determine the offset ($\Vec{d_v}$). Gain factors differ in five levels from 0.8 to 1.2, with a step size of 0.1, indicating 80\% to 120\% of the distance traveled by the real hand. Therefore, with a gain factor smaller/larger than 1, the offset value is smaller/larger than the actual distance traveled.
\[ \vec{d}_{v} = \textit{g} \cdot \vec{d}_{r} \]
The remapped virtual hand position ($\Vec{p_v}$) is determined when the offset ($\Vec{d_v}$) is added from the origin point ($\vec{o}$). The virtual hand that moved the same as the physical hand is remapped according to the offset value from the origin point, moving shorter or longer distances compared to that of the physical hand.
\[ \vec{p}_{v} = \vec{o} + \vec{d}_{v} \]
In short, a gain factor is applied to the distance between the physical hand and the origin point to determine the offset. The offset is then applied to the virtual hand, resulting in a remapped movement. With a gain factor of 1, the virtual hand moves identically to the physical hand, and with a gain factor smaller/larger than 1, the virtual hand moves slower/faster than the real hand. Therefore, while a physical hand moves a certain distance during a task, a remapped virtual hand moves less/more depending on the applied gain factor.


\subsection{Task}
\begin{figure*}[h]
    \vspace{1ex}
    \centerline{\includegraphics[width=\linewidth]{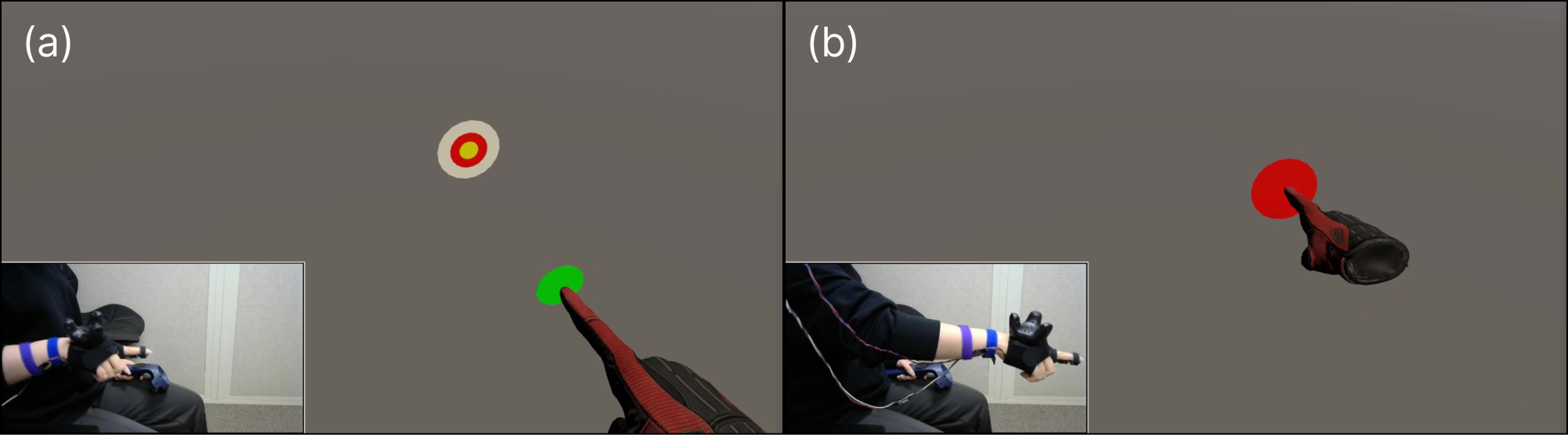}}
    \caption[Task overview]{A participant's first-person view in VR and third-person view of a participant during the hand-reaching task. (a) Participants started a trial by touching the green-colored origin point located near their torso with their right index finger. (b) Participants reached out their right hand to touch the target within an accessible area, and the target turned red when touched.
    } \label{task}
\end{figure*}

The experimental task involved a simple hand-reaching movement toward a target (see Figure~\ref{task}) on each trial. Participants were instructed to make a pointing forward gesture with their index finger and reach out their virtual hand from an origin point to the target as accurately and straight as possible. The green-colored origin point and the target were 30 cm apart in depth in a virtual environment. While the distance the virtual hand moved (origin point to the target) in the virtual environment stayed the same (30cm), the real hand moved shorter or longer distances based on the gain factor: 36cm, 33cm, 30cm, 27cm, and 24cm for the gain factor from 0.8 to 1.2.

As soon as participants touched the origin point, it disappeared from the virtual environment, and both sensory feedback and a gain factor were applied. The target turned red after being touched and disappeared 500 ms later. Then, a user interface panel asking a 2AFC question popped up in front of the field of view, and participants selected one of the two options by pressing the trigger button of the left controller. Since it was important for participants to feel the continuously increasing sensory feedback during the reaching movement, the phrase ``Caution! Too Fast" was displayed after responding to a 2AFC question if the task completion time was too short (less than 300 ms).

For each of the five experimental conditions, participants performed 100 trials consisting of 20 repetitions of the five levels of gain factors. The order of the experimental conditions was counterbalanced with a balanced Latin Square, and trials were presented in a randomized order. Participants performed 500 trials total, excluding practice trials. At the end of each experimental condition, participants were instructed to fill out a sense of embodiment questionnaire \cite{gonzalez2018avatar}, 7-point Likert scale; strongly disagree (-3) to strongly agree (+3). As the Table~\ref{embodiment-tab}, the questionnaire was divided into three subscales: whether the virtual hand was treated as their own hand (Ownership), whether participants could move the virtual hand as if it was their own (Agency), and whether the virtual hand was co-located with their actual hand (Location).

\subsection{Procedure}
All participants first filled out a brief questionnaire providing demographic information and their prior experience in VR and signed the informed consent form. The experimenter explained the overview of the experimental procedure with a prerecorded video. The video showed the appropriate and inappropriate ways to complete the task and the way to answer a 2AFC question. Participants then entered a soundproof room and sat on a chair. With all the equipment attached to the participant's right hand, detailed instructions were given. Participants were instructed to keep the direction of the right hand in line with their right arm and not to bend their wrist, to prevent physical pain from repetitive hand-reaching movements. Participants were also instructed to reach out to the target at a comfortable speed and maintain their initial speed throughout the experiment. After instruction, participants wore VR HMD and adjusted their position and height to level with the origin point.

Before starting each experimental condition in the main experiment, there was a practice phase that consisted of three trials for each condition block (15 practice trials total; three repetitions for each gain factor). After practice trials, participants conducted the task and answered the questionnaire. Then, participants took a 5-minute break and repositioned themselves before the practice phase of the next condition. The duration of the entire experiment was 70 to 80 minutes.

\subsection{Implementation and Setup}
The experiment was conducted in a soundproof room. Participants performed the experimental task with a VR HMD, HTC VIVE Pro Eye (2880 $\times$ 1600 resolution, 110 degrees field of view, 90 Hz refresh rate), in a sitting position. The virtual environment was developed with Unity 3D (2019.4.17f1) and SteamVR Plugin (SDK 1.14.15). We used VIVE Tracker to get the position of participants' right hand and used \cite{zenner2019estimating} gain warp method to implement the retargeting effect. The gain warp redirection method decreases or increases the movement speed of the virtual hand according to the real hand. Participants held a VR controller in their left hand for 2AFC responses, which was visible only in the 2AFC question period. 

\begin{figure}[h]
    \vspace{1ex}
    \centerline{\includegraphics[width=0.55\linewidth]{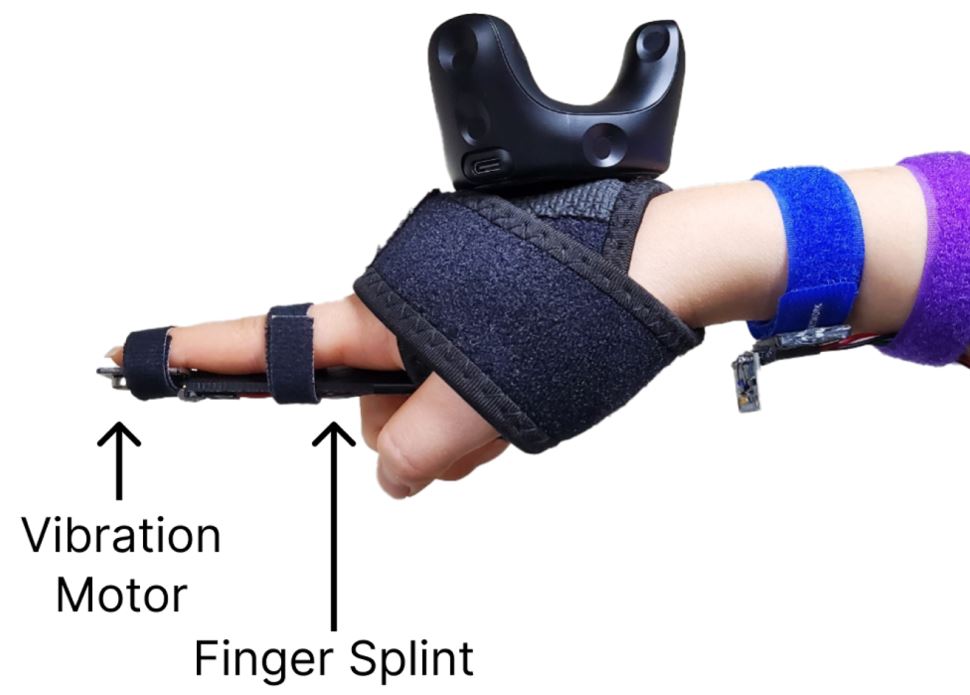}}
    \caption[Equipment setup]{A participant's right hand with all sensors attached. The vibration motor is fixed under the tip of the right index finger; the finger splint helps participants maintain a pointing gesture. The virtual hand moves according to the real hand's movement by the tracker on the back of the participant's hand.
    } \label{righthand}
\end{figure}

As shown in Figure~\ref{righthand}, all the sensors and actuators were tightly fixed on the participants' right hand. A vibration motor (ELB060416) for tactile feedback was placed on the tip of the right index finger and tied with a bundling strap. We provided a 3D-printed finger splint to help participants maintain a pointing gesture with their index finger pointing forward. Holding the finger splint, the VIVE Tracker was wrapped around the back of the participant's hand. Also, we used Raspberry Pi Zero to communicate with the Unity program through TCP (Transmission Control Protocol; Used Python library) to trigger vibration at the right time. The jumper cables that connect sensors with Raspberry Pi were fixed on the ceiling to minimize the extra weight of the equipment on participants' hands. Lastly, we used the VR HMD's built-in speaker for auditory feedback. 

\subsection{Participants}
A total of 30 participants were recruited from a university, and each was compensated approximately \$15 for participation. All the experimental conditions and procedures were approved by the university's Institutional Review Board (IRB), and each participant provided written informed consent. We excluded one participant who failed to differentiate gain factors applied to the virtual hand and derived invalid psychometric functions. Therefore, 29 participants (14 males) aged between 18 and 29 (\textit{M}=24.9; \textit{SD}=2.82) were included in the analysis. Among the participants, two participants never used virtual reality (6.9\%), 12 participants had used it less than five times (41.38\%), 7 participants had used it between 5-10 times (24.14\%), and 8 participants had used it more than ten times (27.59\%).

\subsection{Measures}
In this study, we employed a 2AFC design (Choice between ``Faster" or ``Slower") with the method of constant stimuli \cite{goldstein2016sensation, simpson1988method}. Identifying the percentage of ``Faster" responses on each gain factor derived a psychometric function that defined the PSE (50\% "Faster responses) and upper and lower thresholds (75\% and 25\% of the ``Faster” responses) of remapping detection. The offset of the PSE from the standard point (gain factor of 1) was calculated to analyze perceptual accuracy, with a smaller offset indicating more accurate (veridical) perception. To analyze perceptual sensitivity, the threshold range was calculated as the difference between the upper and lower thresholds, with a narrower range indicating higher perceptual sensitivity. For task performance, we measured the completion time and calculated the precision error of each trial. The precision error indicates the distance from the center of the target to the actual touch point, and thus smaller values represent higher precision. Lastly, the responses to the sense of embodiment questionnaire were collected and analyzed in each subscale.

\begin{table*}[tb]
  \caption{The sense of embodiment questionnaire used in the study. Subsets of questions were chosen from the avatar embodiment questionnaire proposed by \cite{gonzalez2018avatar}. Questions in italics are the control questions. Each question was asked on a 7-point Likert scale (strongly disagree (-3) to strongly agree (+3)).}
  \label{embodiment-tab}
  \scriptsize%
  \centering%
\begin{tabular}{@{}lllll@{}}
\toprule
\textbf{Subscale}  & \textbf{Question}                                                                    &  &  &  \\ \midrule
\textbf{Ownership} & 1. I felt as if the virtual hand I saw was my hand.                                  &  &  &  \\
  & \textit{2. I felt as if the virtual hand I saw was someone else's.}                  &  &  &  \\
          & \textit{3. It seemed as if I might have more than one hand.}                         &  &  &  \\
\textbf{Agency}    & 4. It felt like I could control the virtual hand as if it was my own hand.           &  &  &  \\
          & 5. The movements of the virtual hand were caused by my movements.                    &  &  &  \\
        & 6. I felt as if the movements of the virtual hand were influencing my own movements. &  &  &  \\
       & \textit{7. I felt as if the virtual hand was moving by itself.}                      &  &  &  \\
\textbf{Location}  & 8. I felt as if my hand was located where I saw the virtual hand.                    &  &  &  \\
        & 9. I felt out of my body.                                                            &  &  &  \\ \bottomrule
\end{tabular}
\end{table*}


\subsection{Statistical Analysis}
First, we removed outliers before conducting statistical analyses to exclude extreme responses. The outliers that were more than two standard deviations away from the mean of each participant/each gain factor condition were removed, which was 4.47\% of the total number of trials.

In the statistical analyses, we performed one-/two-way repeated-measures ANOVAs to examine the effects of multisensory feedback and gain factors. If the assumption of sphericity was violated, the Greenhouse-Geisser corrected values were reported and were denoted by $F$\textsubscript{c}. When significant main or interaction effects were found, we conducted two-tailed paired t-tests for post-hoc comparisons. However, when the normality assumption by the Shapiro-Wilk test was violated, Wilcoxon signed-rank tests were conducted instead of paired t-tests. For multisensory feedback, we focused on comparisons between five pairs to investigate the effect of the presence of visual feedback (i.e., \textit{None - V}), the effect of adding auditory or tactile feedback (i.e., \textit{V-VA} and \textit{V-VT}), the integrated effect of adding auditory and tactile feedback (i.e., \textit{V-VAT}), and the differential effects of adding the auditory and tactile feedback (i.e., \textit{VA-VT}). The effect of applied gain factors was analyzed with all ten pairs, but we focused on neighboring gain factors in the interpretation. Post-hoc comparisons were reported with Bonferroni-corrected p-values, and Cohen's \textit{d} and the matched rank biserial correlation (\textit{rB}) were reported as effect sizes. \cite{JASP2022} (version 0.16.2) was used for all statistical analyses with a confidence level of 95\%.

We conducted one-way repeated-measures ANOVAs for the perceptual features (i.e., PSE, upper and lower thresholds, and threshold range) to investigate the effect of multisensory feedback. Two-way repeated-measures ANOVAs were performed for the task performance data. The results of the embodiment questionnaire were nonparametric data, and thus Friedman tests were used in the analysis, and the chi-square and Kendall's \textit{W} values were reported.

\begin{figure*}
    \centerline{\includegraphics[width=\linewidth]{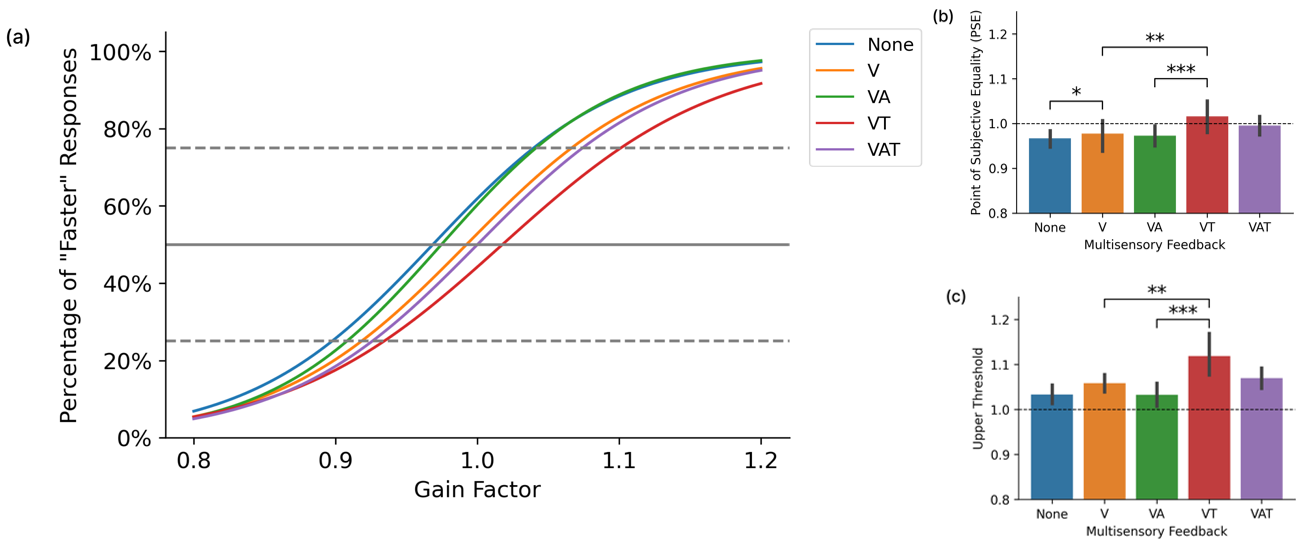}}
    \caption[Psychometric functions and detection thresholds]{The result of perceptual features. (a) Psychometric functions derived from participants' 2AFC responses for each multisensory feedback condition. The x-coordinates for the intersecting points of the psychometric functions and the solid line are the PSEs (50\%) and the two dashed lines are the upper (75\%) and lower (25\%) thresholds. (b) The PSEs and (c) the upper thresholds derived from the psychometric function for each multisensory feedback condition. The PSE and the upper threshold showed analogous statistical differences between conditions, considering the marginally significant difference (\textit{p}=.051) between the \textit{None} and \textit{V} conditions in the upper threshold. The dashed lines represent the gain factor of 1, where the VR hand moved identically to the real hand. The error bars represent 95\% confidence intervals. *\textit{p}$<$.05, **\textit{p}$<$.01, ***\textit{p}$<$.001.
    } \label{2afc}
\end{figure*}

\section{Results}
\subsection{Perceptual Features}
We derived psychometric functions from the 2AFC responses and determined the PSE and the upper and lower thresholds, at which the percentage of ``faster" responses reached 50\%, 75\%, and 25\%, respectively (see Figure~\ref{2afc}). These psychophysical features reflect how each multisensory feedback condition affects the perception of a retargeted hand. The offset of the PSE from the standard point (gain factor of 1) indicates perceptual accuracy to remapping; the smaller the offset, the more accurate (veridical) perception. The range between the upper and lower thresholds, or the threshold range, indicates the interval where people are uncertain or tolerable to changes in gain factor, thus reflecting perceptual sensitivity; the narrower the threshold range, the more sensitive to the changes in the gain factor.

\begin{figure}[h]
    \centerline{\includegraphics[width=0.55\linewidth]{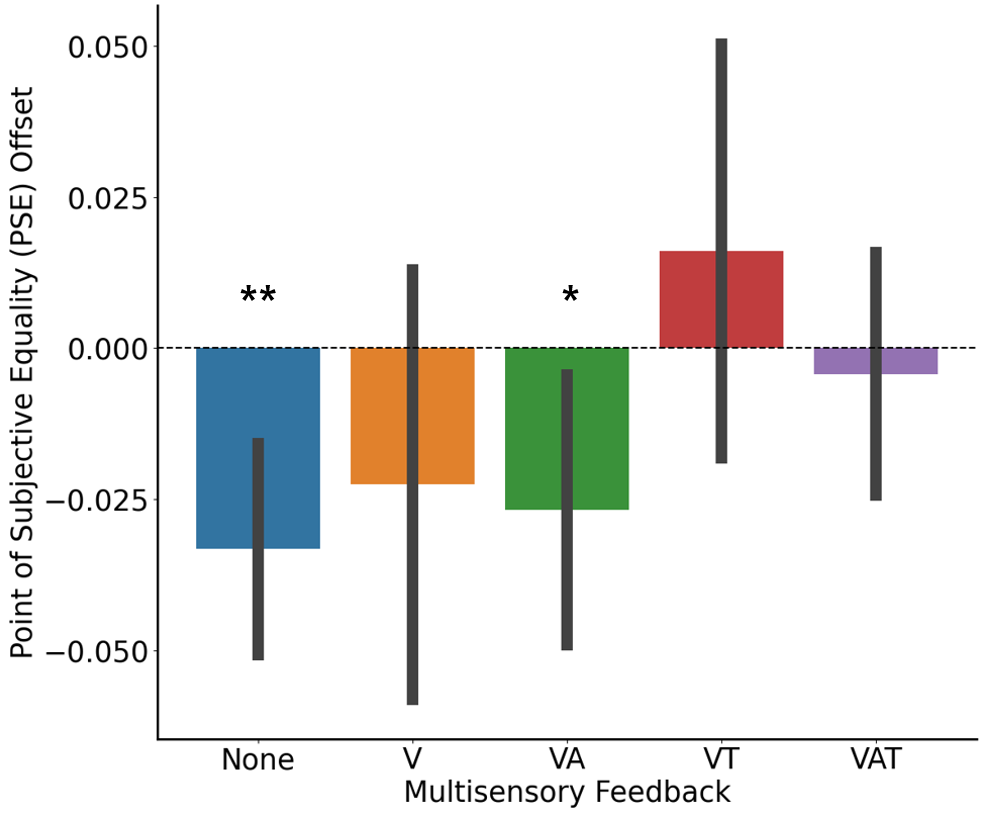}}
    \caption[PSE]{The offset of point of subjective equality (PSE) from the standard gain factor (1.0) for each multisensory feedback condition. The error bars represent 95\% confidence intervals.  *\textit{p}$<$.05, **\textit{p}$<$.01, ***\textit{p}$<$.001.} \label{PSE}
\end{figure}

\subsubsection{Point of Subjective Equality (PSE)}
The main effect of multisensory feedback was statistically significant on the PSE ($F$\textsubscript{c(2.92, 81.86)}=4.39, \textit{p}=.007, $\eta_{p}^{2}$=.136). Bonferroni corrected post-hoc comparisons showed that the \textit{V} condition had a significantly higher PSE ($U$\textsubscript{(28)}=97, \textit{z}=-2.61, \textit{p}=.040, \textit{rB}=-.554) than the \textit{None} condition. The \textit{VT} condition also showed significant differences in the PSEs with the conditions without tactile feedback: higher PSE than the \textit{V} ($U$\textsubscript{(28)}=75, \textit{z}=-3.08, \textit{p}=.007, \textit{rB}=-.655) and \textit{VA} ($U$\textsubscript{(28)}=42, \textit{z}=-3.79, \textit{p}$<$.001, \textit{rB}=-.807) conditions. The PSE of the \textit{None} condition was the smallest among all conditions. The \textit{V} condition did not show any statistical differences from the \textit{VA} and \textit{VAT} conditions ($U\textsubscript{(28)s} \leq 287, \textit{zs} \leq 1.50, \textit{ps} \geq .687, \textit{rBs} \leq .320$).

Next, we conducted single-sample t-tests to evaluate whether the offset of the PSE from the standard gain factor (1.0) was significantly different from zero (veridical perception) for each multisensory feedback condition (see Figure~\ref{PSE}). The mean PSE offset of the \textit{None} (\textit{Mdn}=-.033; \textit{SD}=.050) and \textit{VA} (\textit{M}=-.027; \textit{SD}=.064) was significantly lower than zero. This indicates that under the \textit{None} ($U$\textsubscript{(28)}=84.00, \textit{p}=.004, \textit{rB}=-.66) and \textit{VA} (\textit{t}\textsubscript{(28)}=-2.26, \textit{p}=.032, \textit{d}=-.42) conditions, participants' 2AFC responses were shifted towards the "Faster" response, reflecting the inaccurate perception of the slower movements of the virtual hand as equal to the movement of the physical hand. However, the PSE offset of other feedback conditions (i.e., the \textit{V} (\textit{Mdn}=.003; \textit{SD}=.100), \textit{VT} (\textit{Mdn}=.016; \textit{SD}=.097), and \textit{VAT} (\textit{Mdn}=.012; \textit{SD}=.058)) were not significantly different from zero ($U\textsubscript{(28)s} \leq 289, \textit{ps} \geq .126, \textit{rBs} \leq .33$), indicating that the continuous multisensory feedback was effective in increasing users' perceptual accuracy.  The result that only the conditions that combined the visual or tactile feedback (\textit{V}, \textit{VT}, and \textit{VAT}) achieved high perceptual accuracy suggests that the auditory feedback used in our study did not assist in accurate perception of hand retargeting.

\subsubsection{Upper Threshold}
The upper threshold where participants' ``faster" responses reached 75\% also showed a significant main effect of multisensory feedback ($F$\textsubscript{c(2, 56)}=7.60, \textit{p}=.001, $\eta_{p}^{2}$=.214). Post-hoc comparison results were similar to that of the PSE, revealing that the \textit{VT} condition had a significantly higher upper threshold than the \textit{V} ($U$\textsubscript{(28)}=75, \textit{z}=-3.08, \textit{p}=.007, \textit{rB}=-.655) and \textit{VA} condition ($U$\textsubscript{(28)}=25, \textit{z}=-4.16, \textit{p}$<$.001, \textit{rB}=-.885). The \textit{V} condition also showed no statistical differences between the \textit{VA} and \textit{VAT} conditions ($t\textsubscript{(28)s} \leq 2.40, \textit{ps} \geq .117, \textit{ds} \leq .445$). The \textit{V} condition had a marginally higher upper threshold (\textit{t}\textsubscript{(28)}=-2.75, \textit{p}=.051, \textit{d}=-.511) than the \textit{None} condition.

\subsubsection{Lower Threshold}
The lower threshold, where participants responded ``faster" for 25\%, showed no significant main effect of multisensory feedback ($F$\textsubscript{c(1.33, 37.27)}=.40, \textit{p}=.591, $\eta_{p}^{2}$=.014). The results on the three detection thresholds (PSE, upper, and lower thresholds) on the psychometric function suggest that adding continuous multisensory feedback has more impact on the perception of retargeting at higher gain factors where the virtual hand moves equal to or faster than the real hand.

\subsubsection{Threshold Range}
To examine the effect of multisensory feedback on users' perceptual sensitivity to retargeting, we conducted one-way repeated-measures ANOVA on the threshold range data. Despite the observed statistical differences on the upper threshold, however, the main effect of multisensory feedback on the threshold range was not significant ($F$\textsubscript{c(1.21, 33.77)}=2.12, \textit{p}=.151, $\eta_{p}^{2}$=.070). Combined with the results of PSE offset, this suggests that providing concurrent multisensory feedback can make the perception of hand retargeting more accurate without changing the perceptual sensitivity. 

\subsection{Task Performance}
\subsubsection{Precision Error}
As participants were instructed to touch the center of the target as accurately as possible, task performance was first analyzed by the precision error. The precision error indicates the distance from the point of participants' contact with the target and the center of the target; thus, a smaller precision error represents more accurate performance. The main effect of gain factors on the precision error was statistically significant ($F$\textsubscript{c(2.01, 56.16)}=73.68, \textit{p}$<$.001, $\eta_{p}^{2}$=.725), and all ten pairwise post-hoc comparisons showed significance ($t\textsubscript{(28)s} \leq -4.08, \textit{ps} \leq .003, \textit{ds} \leq -.758 $ or $ \textit{U}\textsubscript{(28)s} \geq 1, \textit{zs} \leq -4.38, \textit{ps}<.001, \textit{rBs} \leq -.931$). The post-hoc comparisons revealed that the higher the gain factor, the larger the precision error; fast movements with high gain factors are relatively difficult to control, resulting in greater errors. The smallest difference between the gain factors was in the 0.8-0.9 pair (\textit{t}\textsubscript{(28)}=-4.08, \textit{p}=.003, \textit{d}=-.758).

The main effect of the multisensory feedback on precision error was not significant ($F$\textsubscript{c(2.49, 69.60)}=1.05, \textit{p}=.368, $\eta_{p}^{2}$=.036), but there was a significant interaction between the gain factors and the multisensory feedback conditions ($F$\textsubscript{c(7.64, 213.78)}=2.12, \textit{p}=.038, $\eta_{p}^{2}$=.070). We conducted post-hoc pairwise comparisons for each gain factor within each feedback condition to determine how gain factors affected behavioral patterns differently in each multisensory feedback condition.

\begin{figure*}[h]
    \vspace{2ex}
    \centerline{\includegraphics[width=\linewidth]{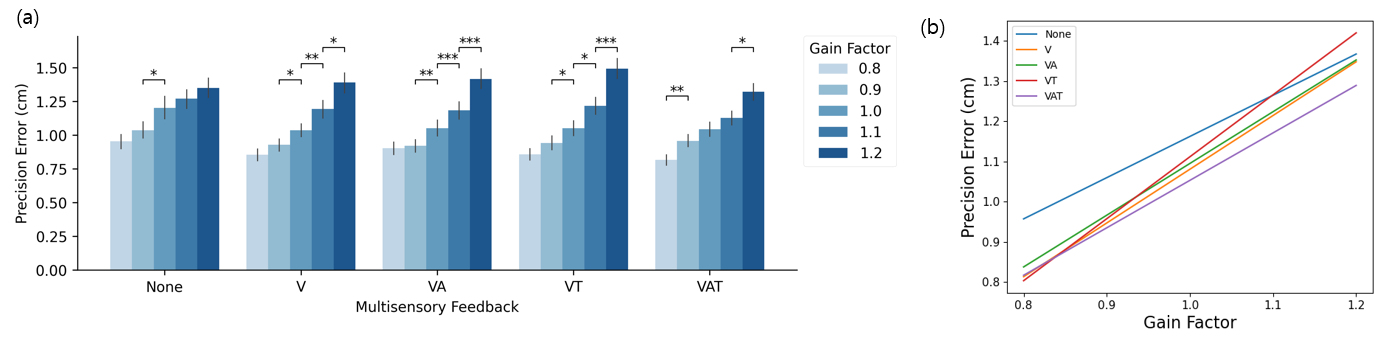}}
    \caption[Precision error]{The analysis result of Precision Error. (a) Precision errors of each gain factor according to the multisensory feedback conditions. The effect of the gain factor on precision error was more prominent when continuous multisensory feedback was given, compared to the None condition. The error bars represent 95\% confidence intervals. *\textit{p}$<$.05, **\textit{p}$<$.01, ***\textit{p}$<$.001. (b) A graph showing the result of fitting precision error values on the linear function to identify the trend according to changes in the gain factor.
    } \label{precision}
\end{figure*}

As shown in Figure~\ref{precision} (a), the \textit{None} condition had the least number of pairs that survived Bonferroni correction. Only the 0.9-1.0 pair showed a statistical difference among the neighboring gain factors ($U$\textsubscript{(28)}=89, \textit{z}=-2.78, \textit{p}=.044, \textit{rB}=-.591). These results indicate that the increasing trend of precision errors for higher gain factors was relatively weak when no sensory feedback was given to movements. Meanwhile, the \textit{V}, \textit{VA}, and \textit{VT} conditions failed to show statistical significance only in the 0.8-0.9 pair. The \textit{VAT} condition, on the other hand, did not show any significant difference in both the 0.9-1.0 and 1.0-1.1 pairs, but it was the only condition that showed a statistically significant difference in the 0.8-0.9 pair ($t$\textsubscript{(28)}=-4.25, \textit{p}=.002, \textit{d}=-.790). In sum, the effect of the gain factor on precision error was more prominent when continuous multisensory feedback was given compared to the \textit{None} condition. This is also illustrated in Figure~\ref{precision} (b), where precision error values are fitted to the linear function, with steeper slopes for conditions with multisensory feedback. Considering that the size of the precision error averaged across gain factors was not different between multisensory feedback conditions, it suggests that getting multisensory feedback assisted accurate performance at lower gain factors, but not at higher gain factors.  



\subsubsection{Task Completion Time}
Task completion time was measured from touching the origin point to touching the target. The main effect of gain factors was significant ($F$\textsubscript{c(1.22, 34.11)}=243.94, \textit{p}$<$.001, $\eta_{p}^{2}$=.897), and post-hoc comparisons proved that differences between all pairs were significant, revealing that task completion time gets shorter as the gain factor increases ($t\textsubscript{(28)s} \geq 11.94, \textit{ps}<.001, \textit{ds} \geq 2.216 $ or $ \textit{U}\textsubscript{(28)s}=435, \textit{zs}=4.70, \textit{ps}<.001, \textit{rBs}=1$). This result was expected because the real hand's traveling distance for each gain factor differed: shorter traveling distances for higher gain factors. On the other hand, the main effect of multisensory feedback was not significant (\textit{F}\textsubscript{c(2.47, 69.04)}=.51, \textit{p}=.638, $\eta_{p}^{2}$=.018), and the interaction between the two factors also did not reach significance ($F$\textsubscript{c(5.65, 158.11)}=.67, \textit{p}=.664, $\eta_{p}^{2}$=.023). These results indicate that the efficiency of task performance was not affected by multisensory feedback conditions.

\begin{figure}[h]
    \centerline{\includegraphics[width=0.55\linewidth]{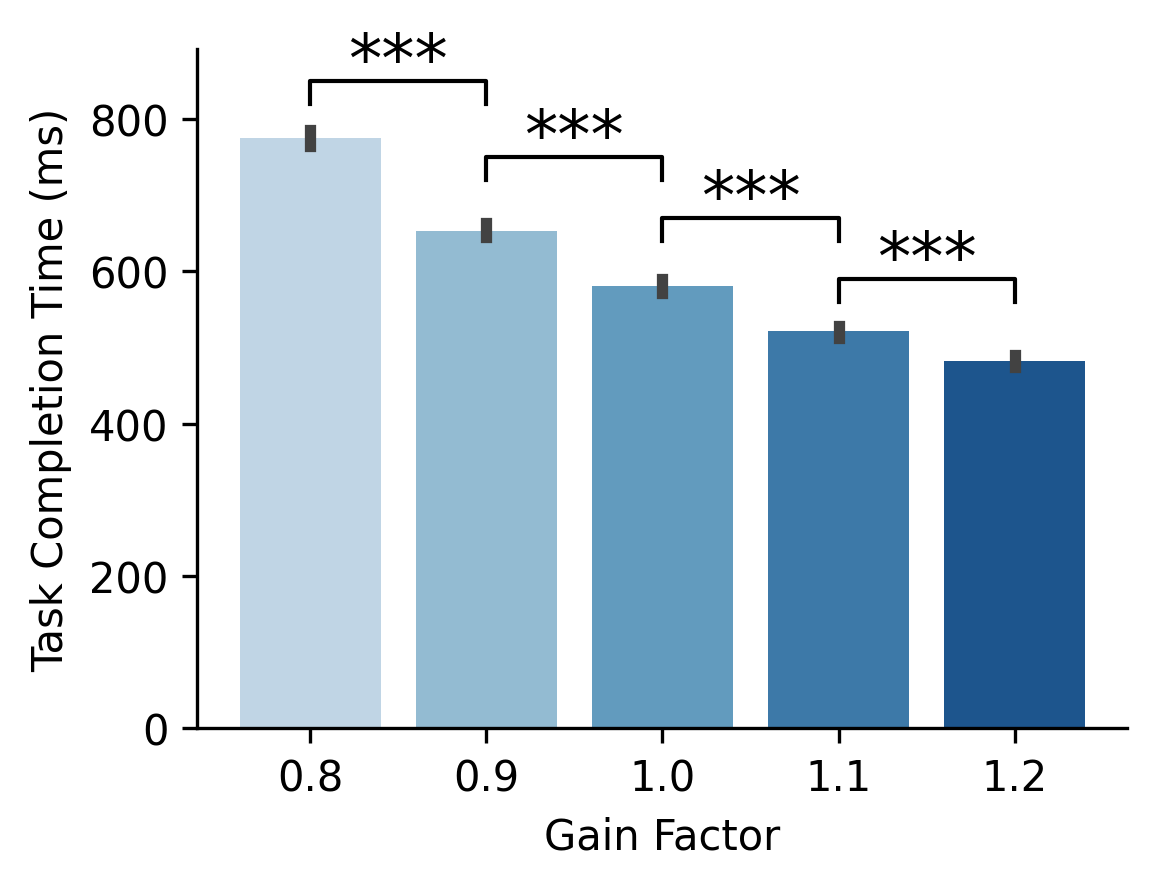}}
    \caption[Task completion time]{Task completion time for each gain factor. All post-hoc comparisons were statistically significant (***\textit{p}$<$.001). The error bars represent 95\% confidence intervals.
    } \label{trialtime}
\end{figure}

\subsection{Embodiment Questionnaire}
We verified the effect of multisensory feedback on embodiment for each subscale of the questionnaire: body ownership, agency, and location. The main effect of multisensory feedback did not reach significance for ownership ($\chi^2$\textsubscript{(4)}=5.08, \textit{p}=.279, \textit{W}=.044), agency ($\chi^2$\textsubscript{(4)}=6.63, \textit{p}=.157, \textit{W}=.057), and location ($\chi^2$\textsubscript{(4)}=5.15, \textit{p}=.272, \textit{W}=.044). This indicates that presenting continuous multisensory feedback, proportional to the distance between the virtual hand and the target, did not affect users' sense of embodiment towards the virtual hand. Within the possible range of scores (-3 to +3), the median scores for all subscales were located closer to the high end across multisensory feedback conditions (see Table \ref{mean-tab}), suggesting that the retargeting parameters and the type of multisensory feedback used in this study were acceptable to participants and did not hinder embodying the virtual hand.

\begin{figure}[h]
    \centerline{\includegraphics[width=0.55\linewidth]{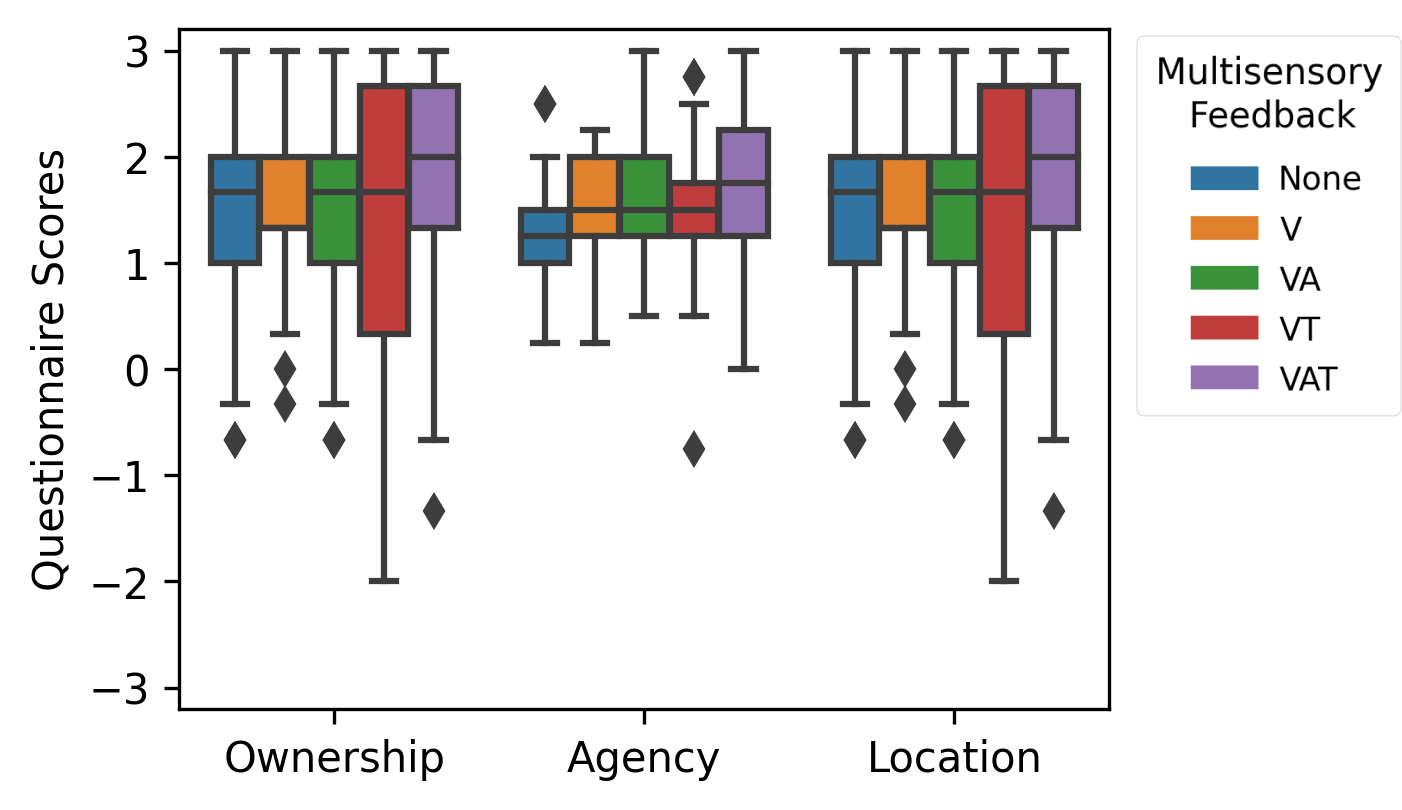}}
    \caption[Questionnaire ratings]{Boxplots of the embodiment questionnaire scores are depicted based on the multisensory feedback conditions and the subscales of avatar embodiment: ownership, agency, and location. The horizontal line above the rectangle indicates the maximum, the line below the rectangle indicates the minimum, and the vertical lines are the whiskers. The top and bottom of the rectangle indicate the third and first quartiles and the line inside the rectangle indicates the median. Markers outside this range indicate outliers.
    } \label{SOE}
\end{figure}

\begin{table*}[tb]
  \caption{The median scores of each subscale of the embodiment questionnaire. Within the possible range of scores (-3 to +3), the median scores of all subscales were rated high across experimental conditions.}
  \label{mean-tab}
  \scriptsize%
  \centering%
\begin{tabular}{@{}llllll@{}}
\toprule
          & None & V    & VA   & VT   & VAT  \\ \midrule
Ownership & 1.67 & 1.33 & 1.67 & 1.67 & 2.00 \\
Agency    & 1.25 & 1.50 & 1.50 & 1.50 & 1.75 \\
Location  & 2.00 & 2.00 & 2.00 & 2.00 & 2.00 \\ \bottomrule
\end{tabular}
\end{table*}

\section{Discussion}
In this study, we specified the perceptual features (psychometric function, PSE, and threshold range) for VR hand-retargeted interaction and verified how users' perception and task performance are influenced by concurrent multisensory feedback. From the experimental results, we found that providing continuous multisensory feedback proportional to the distance between the virtual hand and the reaching destination makes users' perception of hand retargeting more accurate without changing the perceptual sensitivity. Moreover, getting multisensory feedback assisted task performance to be more accurate, prominently at lower gain factors. In this section, we discuss the main results in detail, provide insights for designing future hand-retargeted interactions, and suggest directions for future work based on the limitations of our study.

\subsection{Effect of Multisensory Feedback on Perceptual Accuracy and Sensitivity to Hand Retargeting}
Unlike previous studies that focused only on the PSE ~\cite{burns2006perceptual, lee2015enlarging} to characterize users' perception, we analyzed perceptual accuracy and sensitivity separately by calculating the offset of the PSE from the standard gain factor (1.0) and the threshold range. The results revealed differential effects of multisensory feedback on perceptual accuracy and sensitivity to hand retargeting. 

First, in the baseline (\textit{None}) condition in which no multisensory feedback was given, participants' PSEs were overall shifted towards the gain factors lower than 1.0, indicating that users tend to inaccurately perceive slower movements of the virtual hand as equal to the movement of the physical hand. Getting concurrent multisensory feedback, however, shifted the PSEs closer to the standard gain factor of 1.0, making the perception of hand-retargeting more accurate. Interestingly, the positive effect of multisensory feedback on perceptual accuracy was significant only in those conditions that contained visual or visual combined with tactile feedback (\textit{V}, \textit{VT}, and \textit{VAT}), but not in the condition that combined visual with auditory feedback (\textit{VA}). This suggests that the auditory feedback in this study that changed the pitch continuously according to distance did not help or even hindered accurate perception of hand retargeting. This could be because the association between pitch and distance was unfamiliar, or the range of pitch values used in this study was too narrow to make a significant difference. Nonetheless, we demonstrated that getting continuous visual or tactile feedback proportional to the distance between the virtual hand and the destination effectively increases users' perceptual accuracy in hand retargeting.

Second, perceptual sensitivity to hand retargeting was not affected by concurrent multisensory feedback; the threshold range was not significantly different between multisensory feedback conditions. This result seems contrasting to previous studies that showed that multisensory feedback supported more sensitive target sense perception~\cite{kim2022effect, diaz2006influence, hecht2006multimodal}. However, the contrasting results make sense when considering the difference in the purpose of multisensory feedback; in previous studies, it was designed to augment the target sense itself, whereas in our study, it was to provide additional sensory information consistent with the movement of the virtual hand. Thus, our results actually support the utility of multisensory feedback in designing hand-retargeted interactions; adding continuous multisensory feedback consistent with the virtual movement can make users' perception more accurate without increasing the sensitivity to retargeting. In other words, concurrent multisensory feedback does not decrease the acceptable range of retargeting parameters (threshold range) but still makes the perception more accurate. We also specified the acceptable range of gain factors that can be utilized for hand retargeting for each multisensory feedback condition. For instance, the acceptable range of gain factors for the \textit{VT} condition was between 0.95 and 1.1, with a significantly higher upper threshold value than others.

\subsection{Effect of Multisensory Feedback on Task Performance}
Regarding task performance, we found that as the gain factor applied to the virtual hand increased, the size of the precision error increased, and the task completion time decreased. This pattern of results was expected since higher gain factors mean the faster movement of the virtual hand, which makes precise control of the virtual hand more difficult and the reaching time shorter. Importantly, the effect of the gain factor on precision error was more prominent in conditions with continuous multisensory feedback compared to the \textit{None} condition. As shown in Figure~\ref{precision} (b), the increasing rate of the precision error across gain factors was higher for the conditions with multisensory feedback, reflecting that the concurrent multisensory feedback assisted task performance to be more accurate at lower gain factors, but not at higher gain factors. This tendency was particularly strong in the \textit{VT} condition. These results suggest that providing multisensory feedback, especially visual combined with tactile feedback, in the lower range of gain factors could help users control virtual movements more accurately, but the assisting effect gradually disappears as the gain factor increases.

\subsection{Effect of Multisensory Feedback on Embodiment}
Previous studies have shown that synchronous multisensory stimulation enhances body ownership and agency towards the virtual body \cite{normand2011multisensory, egeberg2016extending}. In our study, on the other hand, concurrent multisensory feedback failed to significantly enhance users' sense of embodiment compared to the baseline (\textit{None}) condition. The reason for the contrasting results could be attributed to the differences in the context of the multisensory feedback. In most previous studies, multisensory feedback consistent with the visual representations in VR was provided to augment the reality of the virtual objects. In our study, the multisensory feedback was consistent with the movement of the virtual hand, while it was \textit{inconsistent} with the real movement as the location of the hand was constantly retargeted. Thus, the constant conflict of information due to the hand-retargeted context may have neutralized the benefit of multisensory feedback on the sense of embodiment. Still, we confirmed that overall scores for the embodiment questionnaire in all conditions were quite high, supporting that the retargeting parameters and the type of multisensory feedback used in this study were acceptable to participants and did not hinder embodying the virtual hand.

\subsection{Implications for Designing Hand-Retargeted Interaction}
We specified users' perceptual accuracy and sensitivity for hand-retargeting and provided acceptable ranges of gain factors using rigorous psychophysical methods. For designing natural and immersive hand-retargeted interactions, it would be critical to use those retargeting parameters such that users cannot reliably notice the inconsistency. Moreover, the results of this study suggest that both perceptual accuracy and the precision of behavioral performance of hand-retargeted interactions vary depending on the concurrently provided sensory feedback. These findings could be utilized to design more effective virtual training applications where trainees need to learn sophisticated skills (e.g., medical robot arm manipulation or automobile maintenance) that require high precision in perception and performance. For instance, practicing under the \textit{VT} condition can more effectively promote skill improvement than other conditions by training to succeed under the most challenging condition, with its highest upper threshold; the higher the upper threshold is, the higher the gain factor can be applied to the virtual hands without trainees' noticing. Repeated practice in controlling the virtual hands under the most challenging conditions would help trainees perform more stably and accurately in real life. In addition, considering the assistive effects of multisensory feedback on the precision of performance at lower gain factors, different combinations of multisensory feedback can be utilized for games involving hand retargeting to design different difficulty levels.

\subsection{Limitations \& Future Work}
There are several limitations of this study. First, this study explored the effect of multisensory feedback when other senses were added centered on visual feedback, due to the limit of experiment trials and the dominant effect of visual stimuli. For future studies, it would be important to explore the different combinations of other sensory modalities and build a cognitive model that represents the weight of each sense on the hand retargeting feedback. Second, we used a simple pointing task as the experimental task to exclude the effects of other potential confounding variables when measuring perceptual and performance characteristics. However, previous studies have reported positive or negative effects of multisensory feedback on the user experience and perception depending on the various contexts and tasks involved~\cite{viciana2010influence, viciana2014influence, weisenberger2004multisensory}. Therefore, it would be essential to expand and apply this feedback system to more complex situations and tasks, such as curve-pointing and grabbing virtual objects. Finally, we relied on a psychophysical method to measure the perceptual characteristics of participants, but it would be nice to complement it with physiological measures, such as electroencephalography (EEG), to provide a more objective measure and comprehensive understanding of the cognitive process~\cite{gehrke2019detecting, gehrke2022neural}.

\section{Conclusion}
In this paper, we provided continuous multisensory feedback during a hand-reaching task to assess its effect on users' perception and performance in hand-retargeted interactions. From the results, we specified the range of retargeting parameters for optimal user experience and performance in VR applications and showed that both perceptual accuracy and the precision of task performance can benefit from concurrently provided multisensory feedback. Based on these findings, we suggest design guidelines and potential applications of hand-retargeted interactions that incorporate the changes in perception and task performance that occur with multisensory feedback. In conclusion, providing multisensory feedback consistent with the virtual movement has the potential to assist users in precise and elaborate control of the virtual body in retargeted interactions, and the degree of the assistance can be manipulated by different combinations of sensory modalities.

\section*{Acknowledgment}
This research is supported by the National Research Council of Science and Technology (NST) (Grant No. CRC 21011), and the National Research Foundation of Korea (NRF) grant (No. 2022R1A4A5033689) funded by the Ministry of Science and ICT (MSIT), Republic of Korea.

\bibliographystyle{unsrtnat}
\bibliography{template}  






\end{document}